\documentclass{src/gistt}

\usepackage{lipsum}
\usepackage{graphicx}
\usepackage{subcaption}
\usepackage[table,xcdraw,dvipsnames]{xcolor}
\usepackage{enumitem}
\usepackage{amsmath}
\usepackage{cleveref}
\usepackage[normalem]{ulem}
\usepackage{textcomp}

\title{Validating Alerts in Cloud-Native Observability}

\author{
Maria C. Borges\\
\texttt{mb@ise.tu-berlin.de}\\
Information Systems Engineering,\\
Technische Universität Berlin
\and
Julian Legler\\
\texttt{jl@ise.tu-berlin.de}\\
Information Systems Engineering,\\
Technische Universität Berlin
\and
Lucca Di Benedetto\\
\texttt{di.benedetto@tu-berlin.de}\\
Technische Universität Berlin
}

\begin{document}

\maketitle

\begin{abstract}
Observability and alerting form the backbone of modern reliability engineering. Alerts help teams catch faults early before they turn into production outages and serve as first clues for troubleshooting. However, designing effective alerts is challenging. They need to strike a fine balance between catching issues early and minimizing false alarms. 
On top of this, alerts often cover uncommon faults, so the code is rarely executed and therefore rarely checked. To address these challenges, several industry practitioners advocate for testing alerting code with the same rigor as application code. Still, there's a lack of tools that support such systematic design and validation of alerts.

This paper introduces a new alerting extension for the observability experimentation tool OXN. It lets engineers experiment with alerts early during development. With OXN, engineers can now tune rules at design time and routinely validate the firing behavior of their alerts, avoiding future problems at runtime.
\end{abstract}

\section{Introduction}
In cloud-native microservice applications, faults are inevitable. When they occur, engineers rely on observability to identify possible service degradations and coordinate a timely response. Alerts in this context act as the smoke alarm, warning teams about faults before they have a chance to cascade into user-facing outages.

The detection mechanism behind alerts is rule-based, using configurable expressions and thresholds. This simplicity means that alerts are relatively easy to set up alongside a monitoring system, especially compared to other ML-based fault detection techniques. However, even though they are easy to set up, designing \textit{effective alerts} can be quite challenging. 

Thresholds must be carefully chosen to reduce false positives, so as not to overwhelm or desensitize engineers, so-called “alert-fatigue” \cite{Google_SREBook_2016}. At the same time, we also want alerts to be timely, to quickly resolve any creeping faults. An alert that is too forgiving may fail to signal a critical issue until it is too late. It's a tricky balancing act, and a challenge that has been well-documented in research \cite{Zhang_MicroserviceSurvey_ThresholdingHard_2019, jagannathan_ibmalertstudy_2023} and industry \cite{Google_SREWorkbook_2018}. 

Another challenge with alert design is that once alerts are configured, validating their operational reliability can be quite difficult. While Prometheus and Cloudfare provide some tools for linting and unit testing alerts, these tests only verify rule syntax and logic and cannot simulate the complex, unpredictable failure patterns of production environments.  

Even though alerting is such a critical aspect of cloud-native engineering, alert design and validation remain surprisingly informal. Design relies predominantly on developer experience or ad-hoc reconfigurations after major incidents \cite{niedermaier_ObservabilityInterviewStudy_2019}, while validation is minimal or non-existent. To ensure better alerting practices moving forward, industry practitioners have begun compiling alert design guidelines \cite{beyer_alertingguide_2019}, while in research, some first promising design approaches, e.g. based on historical alert analysis \cite{jagannathan_ibmalertstudy_2023}, have started to emerge. Despite these  efforts, there is still a general lack of practical tooling to support teams with their alerting practices.

In this paper, we build upon our previous work around observability assessment and experimentation \cite{Borges_InformedObservabilityDecisions_2024, Borges_Method_2025}. We extend our observability experiment tool OXN  \cite{ Borges_OXN_2024} to support alert design and validation. In the following, we revisit OXN's architecture and show the design changes introduced for alert experimentation. To demonstrate its use, we evaluate two different strategies for alert rules and discuss future directions.

\section{OXN}
In order to make informed design decisions around observability and weigh between design alternatives, engineers need a way to measure how effective their observability actually is. 

To support this, we developed the \textbf{Observability eXperiment eNgine (OXN) }\cite{Borges_OXN_2024,Borges_InformedObservabilityDecisions_2024}, a tool that enables such assessments during design time, through controlled experiments\footnote{Docker-compose version: \texttt{\url{github.com/nymphbox/oxn}}}\footnote{Kubernetes version: \texttt{\url{github.com/LHMoritz/oxn-fork}}}. OXN combines fault injection/chaos engineering with observability tuning, allowing experimenters to explore how different observability setups respond to various fault scenarios.

OXN can deploy any cloud-native microservice application as the system under experiment (SUE), using a kubernetes helm chart. After deploying the SUE, OX\textbf{}N starts a workload generator, where load shapes can easily be specified via locust files. Before and during the experiment, OXN applies treatments, which are controlled changes to the system under experiment. We distinguish between fault treatments and observability treatments. With fault treatments, experimenters can inject faults during the operation of the application. With observability treatments, they can change the observability of the application, e.g. by enabling certain instrumentation points, changing collection sampling, or adding new alerting rules. Our previous paper \cite{Borges_OXN_2024} includes a detailed description of OXN's architecture and mode of operation.
\vspace{0.5em}

\noindent \textit{Alerting Extension}

\noindent Previous versions of OXN supported fault detection only through offline methods. The system would run experiments and, upon completion, analyze the collected observability data using fault detection algorithms (e.g. classifiers) to determine whether the injected faults were detectable.

With this alerting extension, we enable measurement of online fault detection, including critical metrics such as time-to-detect. Alerts are specified using prometheus altering rules. The prometheus alerting component is deployed as part of the SUE. During the experiment execution, OXN collects alerts from this component and records their timestamps. These alert timestamps are then compared with the fault injection timestamps to classify detections into true positives, false positives and false negatives. 

While trying to measure alert accuracy, we were confronted with a fundamental challenge: defining what constitutes a valid detection versus a false positive is inherently context-dependent. For instance, should an alert that triggers 30 seconds after a fault injection has ended be considered a false positive?  To accommodate this, our extension provides flexibility for practitioners to define their own conditions for what constitutes valid fault detection. 

For this reason, while we have implemented alert precision and recall as a feature in OXN, we do not include an evaluation based on these metrics in this paper, as it would involve too many arbitrary assumptions. Instead, in the next section, we present an exemplary alert validation that highlights two different dimensions of alert design.
\section{Exemplary Alert Validation}

 In this section, we demonstrate the applicability of our approach for validating alerts. With OXN, practitioners are able to compare the behavior of different alert rules for specific fault scenarios. 
 
For our demonstration, we chose two of the alert design options outlined in the Google SRE Workbook \cite{Google_SREWorkbook_2018} and apply them to a popular opensource cloud-native application.
\vspace{0.5em}

\noindent \textit{SUE Setup and Experiment Specification}

\noindent As our SUE, we use the OpenTelemetry Astronomy Shop Demo\footnote{\url{github.com/open-telemetry/opentelemetry-demo/}} microservice application,
which is a community project intended to illustrate the use
of different observability tools in a near realworld environment. 
The application consists of 20 microservices and is instrumented to collect several metrics, including request and error rate. 

For the experiments, we deploy the application in a cloud-based, two-node kubenetes cluster\footnote{each node with 4vCPUs and 16GB Memory}. We simulate a load of 800 concurrent users, keeping it constant throughout the entire duration.
For our experiments, we focus on the frontend service. We design alerts around the request error rate, a common Service Level Indicator (SLI) in production systems. To calculate the error rate, we use metric instrumentation already present in the application. 

As the `alert-worthy' fault scenario, we employ packet loss injection to simulate service degradation conditions. In our fault scenario, we introduce three intermittent 1min-long phases where packet loss spikes up to 25\%. This simulates common infrastructure issues such as a misconfigured load balancer and autoscaler. We execute this fault pattern several times in succession, with cooldown periods in between, so that we can have several data points for analysis. 

\vspace{0.5em}

\begin{figure*}[t]
    \centering
    \includegraphics[trim={0.3cm, 0.45cm, 0, 0.27cm}, clip, width=0.932\textwidth]{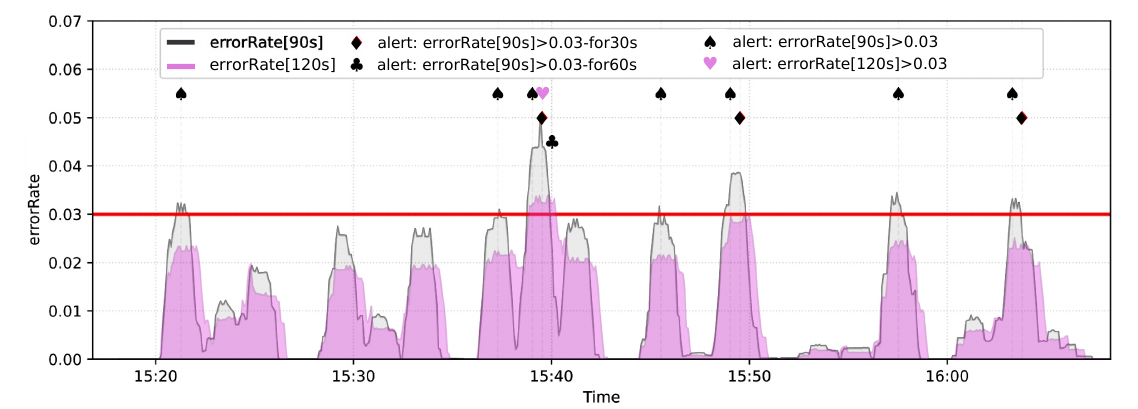}
    \caption{Error rate and alerts triggered during exemplary experiment run}
    \label{fig:plot}
\end{figure*}

\noindent \textit{Alert Design Alternatives}

\noindent As a starting point, we consider an alert that defines a simple and immediate threshold violation:
\vspace{0.3em}

\begin{footnotesize}
\texttt{alert: HighErrorRate}

\texttt{expr: errorRate[90s]\textgreater 0.03}
\end{footnotesize}

\vspace{0.3em}
This alert triggers as soon as the error rate for a 90s window exceeds the defined threshold of 3\%. While this approach offers certain benefits, namely quick detection with low time-to-detect, it also presents disadvantages, particularly being too ``trigger-happy" and potentially generating excessive alerts.

To mitigate these issues, practitioners can pursue two primary strategies. The first is to increase the time window for error rate calculation: 
\vspace{0.3em}

\begin{footnotesize}
\texttt{alert: HighErrorRate}

\texttt{expr: errorRate[120s]\textgreater 0.03}
\end{footnotesize}

\vspace{0.3em}
With this change, the alert becomes less sensitive but maintains the same threshold. It smooths out temporary spikes but increases the time-to-detect. 

The other strategy is to add a duration condition for the threshold of the alert:
\vspace{0.3em}

\begin{footnotesize}
\texttt{alert: HighErrorRate}

\texttt{expr: errorRate[90s]\textgreater 0.03}

\texttt{for: 60s}
\end{footnotesize}

\vspace{0.3em}
This alert maintains the original 90s rate calculation window but requires the condition to persist for 60s before triggering. The advantage of this approach is that alerts require a sustained degradation before firing, which means that alerts are more likely to correspond to a significant event. However, the disadvantage is that if the metric even momentarily returns to a level below the threshold, the duration timer resets and the alert never triggers.
\vspace{0.5em}

\noindent \textit{Experiment-based Alert Comparison}

\noindent With OXN, practitioners can validate these alert design assumptions and test them for their concrete application and observability setup, to arrive at the most suitable alert for their needs. 

Figure \ref{fig:plot} shows an example of such an experiment. 
The baseline alert \texttt{errorRate[90s]\textgreater0.03} triggers very frequently, in five of the six injected patterns, sometimes even triggering multiple times per fault scenario pattern. As expected, increasing the error rate window to 120s smooths out the curve, but now the alert triggers only once, in the most severe service degradation. As for the other strategy, adding a duration condition also decreases the number of alerts, though this behaves more variably as it is influenced by the spikes within the fault scenario pattern.

With this example, we do not aim to answer which alert design is best. Rather, we demonstrate that practitioners can now systematically assess alert behavior through experimentation. Our results for this concrete experiment also confirm that the observed trends align with theoretical predictions.
\section{Conclusion} With OXN, we developed the first tool for experiment-driven observability assessment. This new extension lets practitioners compare different alerting decisions, as well as validate alert triggering behavior. We demonstrated the systematic approach by comparing two different alert design strategies.

While we focused here on alert rules, fault detection also depends on other underlying observability factors, like instrumentation, collection rates, or even the metric data model. OXN can test these in isolation or also together. However, this opens up a large design space that is time-consuming to explore. Through batch experiments, OXN already makes first steps towards automating this observability design space exploration. However, an obvious next step is to develop an intelligent OXN—potentially AI-driven—that can evaluate and recommend observability designs automatically.

\printbibliography

\end{document}